\documentstyle[10pt,aas2pp4]{article}

\lefthead{Kavelaars}
\righthead{Globulars Clusters and Galaxy Formation}
           
\begin{document}
\small

\title{On the relation between globular cluster specific frequency and 
galaxy type.}

\author{JJ Kavelaars}
\affil{Department of Physics and Astronomy, McMaster University}
\authoraddr{Hamilton ON  L8K 2P4}

\begin{abstract}
The universality of the globular cluster luminosity function (GCLF)
contrasts the variation seen in the specific frequency ($S_N$).  The
variation in $S_N$ has been shown to follow a linear relation with
$L_X$ for brightest cluster galaxies \cite{1997ApJ...481L..59B}.
Further, the variation of $S_N$ with galactic radius within individual
giant ellipticals is seen to be a constant fraction of the gas density
\cite{1998JRASC...M}.  There are now a number of galaxies for which
direct mass estimates based on the radial velocities of the globular
clusters are available.  By comparing the mass of galaxies determined in
this way with the number of clusters within these galaxies we show 
that the fraction of mass which is converted into globular clusters is
constant independent of galaxy type or environment.  This implies that
the process of globular cluster formation is not influenced by the
host galaxy and supports the notion of the universal GCLF.
\end{abstract}

\section{Introduction}
Globular clusters (GCs) are ubiquitous and apparently obey a universal
formation mechanism \cite{1994ApJ...429..117H} as evidence by the constancy of
the globlar cluster luminosity function (GCLF)
\cite{1991ARA&A..29..543H}.  However, this constancy of formation
appears juxtaposed against the environments of the GCs host galaxies.
The nature of this juxtaposition manifests itself in the variation of
cluster specific frequency as a function of galaxy type and
environment.

In order to compare globular cluster populations of two disparate galaxies
(say an E galaxy compared to an S0) the total number of clusters about
that galaxy should be normalized to some trait which both objects have
in common.  The common measure in use is referred to
as the globular cluster specific frequency and is defined as 
\begin{equation}
S_N = N_t \time 10^{0.4(M_V^T + 15.0)}
\end{equation}
\cite{1981AJ.....86.1627H}, where $N_t$ is the total cluster
population and $M_V^T$ is the integrated V magnitude of the
galaxy.

Observations of the globular cluster systems of brightest cluster
galaxies (BCGs) have revealed the compelling relationship between GC
specific frequency and the X-ray luminosity ($L_X$) of the host galaxy
\cite{1997ApJ...481L..59B,1998AJ....115.1801H}.  This relation gives
an intuitive explanation of the variation in $S_N$ amongst the
galaxies of the same type and suggests that galaxy mass is the
determining factor in cluster formation.

Use of X-ray temperatures ($T_X$) to determine the radially dependent gas
density within M87 and comparing this to the radial variations in
cluster numbers has further revealed that the efficiency of cluster
formation is a constant of the density of gas
\cite{1998JRASC...M}.  The evidence that the formation
efficiency of globulars is a constant with radius is reassuring to
the previous result for BCGs as the relation between cluster numbers
and $L_X$ was based solely on clusters within a distance $32h^{-1}kpc$ of
the BCG center.

The final piece of the $S_N$ puzzle is the variation
in $S_N$ between spiral and elliptical galaxies.

\section{$S_N$ and galaxy type}

 Use of the $S_N$ relation leads to the common result that spiral
galaxies have a lower specific frequency than do ellipticals
(Fig.~\ref{fig:SN}).  This result is further emphasized when the
high $S_N$ galaxies M87 and NGC 1399 are included and
their total cluster populations are compared to their total
luminosity. The reason for this variation of specific
frequency is not yet well understood although many competing theories have
been proposed and not all of these suggestions are obviously wrong.

The development of multi-object spectrographic devices has made the
determination of the mass of remote galaxies using their cluster
populations possible.  These investigations use the clusters as test
particles in the potential well of the parent galaxies mass
distribution and provide a reasonable measure of the total mass of the
galaxy (for example Bridges et al. 1997 \nocite{1997MNRAS.284..376B}).
Although the full decomposition of the mass distribution is not
possible \cite{1993AJ....106.2229M} the use of a model dependent mass
estimator such as the projected mass estimator (PME)
\cite{1981ApJ...244..805B} makes possible a comparison of the masses
of the parent galaxies using the clusters themselves.

Several studies of galaxy halo masses using GCs as probes are now
underway and the number of published results now makes clear the link
between total cluster population and the mass of the parent galaxy.

\section{Mass-Specific Frequency}

Table~\ref{table:mass} lists those galaxy whose masses have been
determined via spectroscopy of the cluster population.  For each of
these galaxies the mass has been estimated using the PME modified to
account for extended mass distributions \cite{1985ApJ...298....8H}.
These estimates are not necessarily the most accurate for each
individual galaxy (in the cases of M87 \cite{1997ApJ...486..230C} and
NGC 1399 \cite{1998AJ....115..105K} the large numbers of
velocities available make possible a more complex model analysis)
however, by using the same estimator similar biases are introduced for
each galaxy and so a fair comparison of the relative mass of these
galaxies is possible.  Also shown is the radius out to which cluster
velocities have been measured.  When using the PME to determine the
mass of a distribution the mass determined is that enclosed by the
most distant test particles. Thus the mass quoted in the table is in
fact the mass enclosed in the radius given for that galaxy.

\begin{deluxetable}{lclcccll}
\tablecaption{Galaxy Properties\label{table:mass}}
\small
\tablehead{
\colhead{Galaxy} &
\colhead{Type} &
\colhead{$M_V^T$} &
\colhead{Mass} & 
\colhead{$N_t$} &
\colhead{Radius} &
\colhead{$S_N$}&
\colhead{$S_M$} \nl
 & &  &
\colhead{$(10^{11}M_\odot$)} & & \colhead{$kpc$} & & \colhead{$10^{-4}$} }
\startdata
NGC 1399 &  E1/cD & -21.1 & $50\pm10$\tablenotemark{h} & $2100\pm1000$\tablenotemark{a} & 25 
& $19\pm6$ & $1.0\pm0.5$ \nl			       	  
NGC 4486 &  E0/cD & -22.4 & $60\pm20$\tablenotemark{i} & $4800\pm500$\tablenotemark{b}  & 18 
& $14\pm0.5$ & $1.9\pm0.6$ \nl			       	  
NGC 4472 &  E2  & -22.6 & $25\pm5$\tablenotemark{j}    & $2200\pm600$\tablenotemark{c}  & 18 
&  $5.6\pm1.7$ & $2.0\pm0.7$ \nl			       	  
NGC 4594 & Sa   & -22.3  & $5\pm1.5$\tablenotemark{k}  & $660\pm200$\tablenotemark{d}   & 14 
& $2.3\pm0.7$ & $1.9\pm0.8$ \nl			       	  
NGC 3115 & S0   & -21.45 & $10\pm4$\tablenotemark{e}   & $415\pm50$\tablenotemark{e}    & 19 
&$2.0\pm0.5$ & $1.6\pm0.6$ \nl			       	  
NGC 3031 & Sab  & -21.2  & $3\pm1$\tablenotemark{l}    & $210\pm30$\tablenotemark{f}    & 20 
& $0.7\pm0.1$ & $1.5\pm0.6$ \nl			       	  
M31  &  Sb  &  -21.7   & $4\pm0.4$\tablenotemark{m}    & $250\pm100$\tablenotemark{g}  & 20 
&  $0.7\pm0.2$ & $1.5\pm0.9$ \nl

\nocite{1991AJ....101..469B}
\nocite{1994ApJ...422..486M}
\nocite{1981AJ.....86.1627H}
\nocite{1992AJ....103..800B}
\nocite{1998thesis.phd.jjk}
\nocite{1995AJ....109.1055P}
\nocite{1994AJ....107..555R}
\nocite{1998AJ....115..105K}
\nocite{1987AJ.....93..779H}
\nocite{Sharples...Private}
\nocite{1995AJ....110..620P}
\nocite{1993AA....274...87F}
\enddata

\tablecomments{$S_N$ \cite{1991ARA&A..29..543H} determined using the
entire cluster population and full integrated luminosity of the parent
galaxy. The value for the specific mass frequency was determined using
only clusters interior to the radius to which the mass of the galaxy
has been determined. Keys refer to the reference section.}

\end{deluxetable}

As shown in Table~\ref{table:mass} the mass-to-light ratios of these
various galaxies are not constant.  In fact the immediate indication is
that galaxies with high values of $S_N$ are also those with high
values of $M/L$.  This begs the question of the mass-specific
frequency.  Defining a mass-specific frequency as the ratio of a
galaxies cluster population interior to some radius, $R$, to the mass
of the galaxy within that radius provides the relation,
\begin{equation}
S_M = N_T(R) \frac{23.2\times10^{4} M_\odot}{M_{gal}(R)}
\end{equation}
where $N_T(R)$ is the total cluster population interior to some radius
$R$ and $M_{gal}(R)$ is the mass interior to that radius, as
determined via the PME.  The normalization factor ($23.2\times10^4
M_\odot$) is the mean mass of the Milky Way GCs assuming $M/L = 2$.
\cite{1996AJ....112.1487H}.  This definition is different from
previous ones which relied on a constant value of $M/L$ for a given
galaxy type \cite{1991ARA&A..29..543H,1993MNRAS.264..611Z}.

\begin{figure}
\plottwo{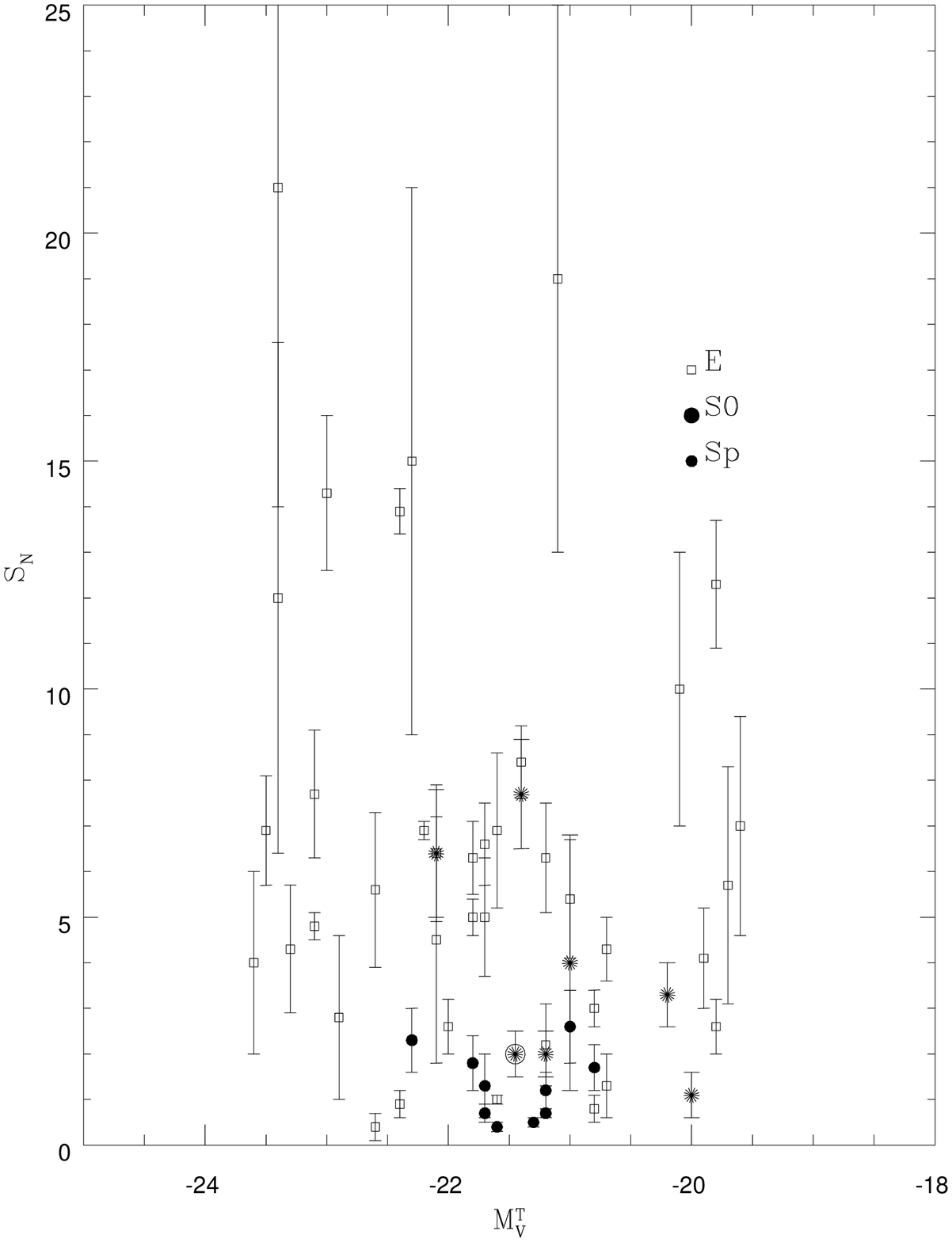}{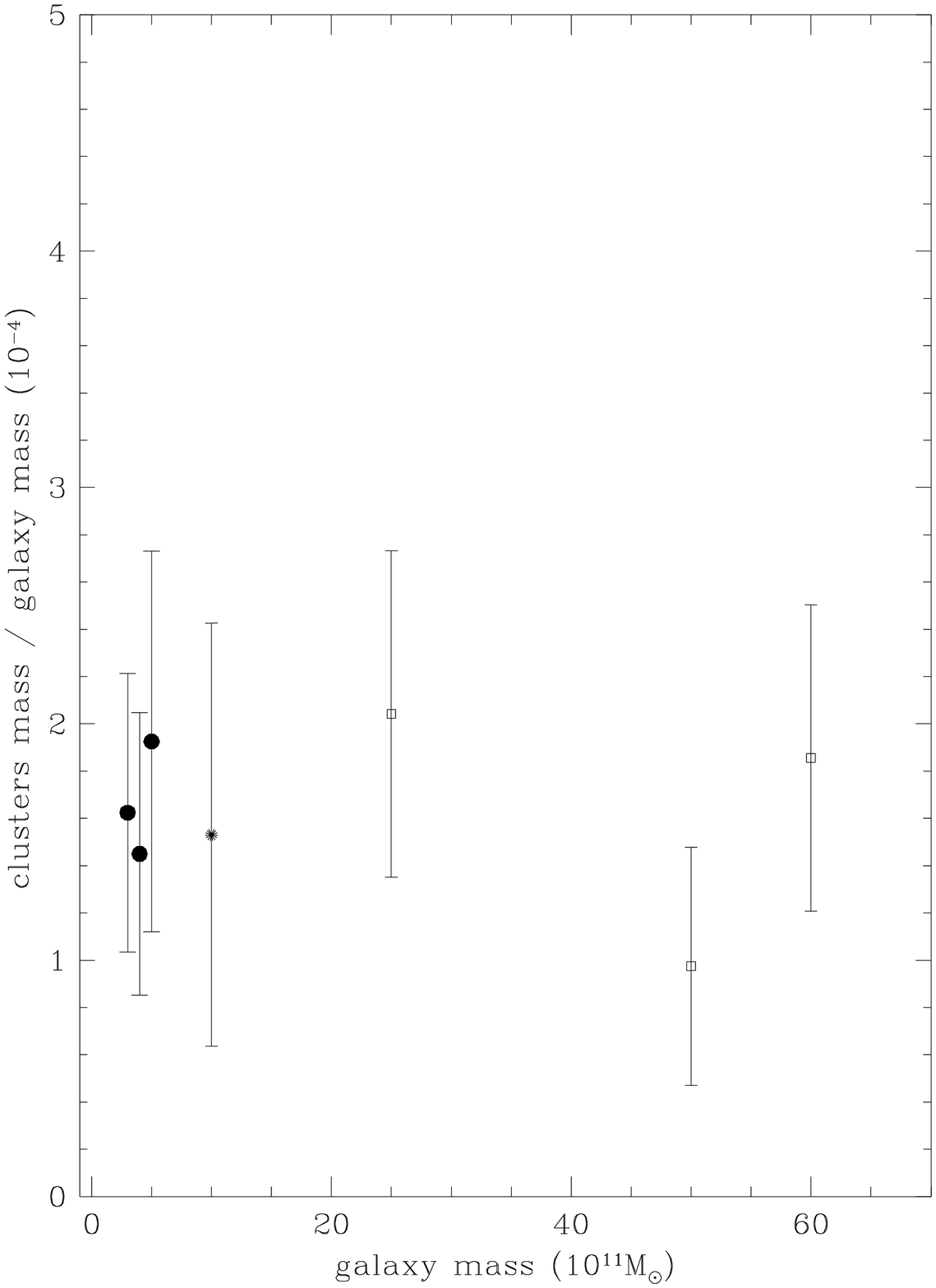}
\caption{a) The specific luminosity frequency of GCs as a function
of parent galaxy luminosity. b) The specific mass frequency of GCs as a
function of parent galaxy mass.\label{fig:SN}\label{fig:mass}}
\end{figure}

The values of $S_M$ are listed in column 6 of
Table~\ref{table:mass} and the relation between parent galaxy
luminosity and $S_M$ is shown in Fig.~\ref{fig:mass}.  Clearly
there is now no correlation between galaxy luminosity and cluster population
and no obvious difference between the relative populations of spirals,
ellipticals and cD galaxies, even those previously seen as
super-abundant can now be reconciled by a single value of cluster
formation efficiency.  


The cause of the anomalous $S_N$ values of some cD galaxies and the
difference between $S_N$ of spirals and that of ellipticals must lie
in the stellar population.  Already in the literature there are some
suggestions to this explanation of the $S_N$ values
\cite{1997ApJ...481L..59B,1998AJ....115.1801H}.  As the efficiency of
cluster formation appears to be universal then it must be the star
formation efficiency which varies.  If the early burst of
star formation results in a galactic wind shortly after the
clusters formed but while there was still considerable gas left
\cite{1998AJ....115.1801H} then a high $S_N$ system could result,
leaving the expelled X-ray gas in the galaxy's halo.  Here the obvious
discriminator between high $S_N$ and low $S_N$ systems is the rate and
density of cluster formation and not its efficiency.  At the time of
formation of the GCs the super giant molecular clouds (postulated to
be the progenitor objcts of GCs \cite{1994ApJ...429..117H}) would provide sheilding against the UV radiation of the first generation of star formation, gas ejected from the progenator would, however, find itself photo-ionized and unable to cool and form stars. 



\section{Implication for galaxy formation}
Accepting the preceding as true results in an implication regarding
the star formation rate of a galaxy.  Those galaxies with low values
of $S_N$ are either more efficient at forming stars than their high
mass cousins or the initial mass function of star formation in dense
galaxies is tilted towards a preference for very massive stars early
in that galaxies history.  Although this first round of massive
objects still contribute their mass they no longer contribute to the
luminosity of the galaxy.  In addition the universal efficiency of
cluster formation implies that cluster formation occurs very early in
the galaxy formation process, prior to the type of galaxy which will
form having any chance to influence the number of cluster which it
will posses.

The notion that a stellar wind might drive gas away from giant E
galaxies could actually be the factor which eventually determines the
type of galaxy which will form.  Those galaxies with lower densities
of initial cluster formation can have their proto-galactic chunks
dissipationally settle into a disk while those which form in dense
regions with large numbers of clusters and a strong galactic wind will
not manage to drag enough gas down into a plane in order that a disk might
form.  

\section{The universality of the GCLF}
As stated in the introduction the universality of the GCLF is
paradoxical when compared to $S_N$, however when compared to $S_M$ the
paradox is removed.  The correlation of galaxy mass with total
population implies that the majority of clusters form prior to the
determination of the final galaxy type.  During this epoch the
potential well of the proto-galaxy is likely to be much softer and the
destruction of GCs via tidal stripping and evaporation is likely to
dominate other modes of destruction.  Further these modes are mostly
dependent on the mass of clusters formed and so the proportion
destroyed is likely to be constant among different proto-galaxies.
The early epoch of cluster formation, when clouds massive
enough to eventually form clusters are present, occurs with uniform
efficiency in all the environments studied to date.  Further measures
of the mass distribution about galaxies of various types, sizes and stages
of evolution should solidify this result.

THe conclusions of this work are:
\begin{itemize}
\item Globular clusters are create with an efficiency which is independent of
galaxy type and enviroment.

\item Given the correlation between cluster numbers and mass and the run of
$S_N$ with galaxy type: elliptical galaxies are less efficient at producing stars than are spirals.

\end{itemize}

\end{document}